\documentclass[prl,aps,superscriptaddress,twocolumn]{revtex4-1}
\setcitestyle{super}
\usepackage{amsfonts,amssymb,amscd,amsthm}
\usepackage{graphicx}
\usepackage{threeparttable}
\usepackage{multirow}
\usepackage{mathrsfs}
\usepackage[intlimits]{amsmath}
\usepackage{array}
\usepackage{xcolor}
\usepackage[colorlinks, citecolor=red]{hyperref}
\usepackage{lineno}
\usepackage{etoolbox}
\usepackage{color,soul}
\usepackage[normalem]{ulem}

\definecolor{blue}{rgb}{0,0,1}
\definecolor{red}{rgb}{1,0,0}
\definecolor{green}{rgb}{0,1,0}

\begin{document}

\title{A scalable edge-pass Purcell filter for high-fidelity readout of superconducting qubits}

\author{Xudong Liao}
\altaffiliation{These three authors contributed equally to this work.}
\affiliation{National Laboratory of Solid State Microstructures, School of Physics, Nanjing University, Nanjing, 210093 Jiangsu, China}
\affiliation{Shishan Laboratory, Nanjing University, Suzhou, 215163 Jiangsu, China}
\affiliation{Jiangsu Key Laboratory of Quantum Information Science and Technology, Nanjing University, Suzhou, 215163 Jiangsu, China}

\author{Yuan Li}
\altaffiliation{These three authors contributed equally to this work.}
\affiliation{Tencent Quantum Laboratory, Tencent, Shenzhen, Guangdong 518057, China}

\author{Sainan Huai}
\altaffiliation{These three authors contributed equally to this work.}
\affiliation{Tencent Quantum Laboratory, Tencent, Shenzhen, Guangdong 518057, China}

\author{Shuyi Pan}
\affiliation{National Laboratory of Solid State Microstructures, School of Physics, Nanjing University, Nanjing, 210093 Jiangsu, China}
\affiliation{Shishan Laboratory, Nanjing University, Suzhou, 215163 Jiangsu, China}
\affiliation{Jiangsu Key Laboratory of Quantum Information Science and Technology, Nanjing University, Suzhou, 215163 Jiangsu, China}

\author{Zhenxing Zhang}
\affiliation{Tencent Quantum Laboratory, Tencent, Shenzhen, Guangdong 518057, China}

\author{Zhiwen Zong}
\affiliation{Tencent Quantum Laboratory, Tencent, Shenzhen, Guangdong 518057, China}

\author{Kunliang Bu}
\affiliation{Tencent Quantum Laboratory, Tencent, Shenzhen, Guangdong 518057, China}

\author{Yulei Ye}
\affiliation{Tencent Quantum Laboratory, Tencent, Shenzhen, Guangdong 518057, China}

\author{Wen Zheng}
\affiliation{National Laboratory of Solid State Microstructures, School of Physics, Nanjing University, Nanjing, 210093 Jiangsu, China}
\affiliation{Shishan Laboratory, Nanjing University, Suzhou, 215163 Jiangsu, China}
\affiliation{Jiangsu Key Laboratory of Quantum Information Science and Technology, Nanjing University, Suzhou, 215163 Jiangsu, China}

\author{Xinsheng Tan}
\affiliation{National Laboratory of Solid State Microstructures, School of Physics, Nanjing University, Nanjing, 210093 Jiangsu, China}
\affiliation{Shishan Laboratory, Nanjing University, Suzhou, 215163 Jiangsu, China}
\affiliation{Jiangsu Key Laboratory of Quantum Information Science and Technology, Nanjing University, Suzhou, 215163 Jiangsu, China}

\author{Yang Yu}
\affiliation{National Laboratory of Solid State Microstructures, School of Physics, Nanjing University, Nanjing, 210093 Jiangsu, China}
\affiliation{Shishan Laboratory, Nanjing University, Suzhou, 215163 Jiangsu, China}
\affiliation{Jiangsu Key Laboratory of Quantum Information Science and Technology, Nanjing University, Suzhou, 215163 Jiangsu, China}

\author{Xiaopei Yang}
\email{xiaopeiyang@tencent.com}
\affiliation{Tencent Quantum Laboratory, Tencent, Shenzhen, Guangdong 518057, China}

\author{Tianqi Cai}
\email{tsefctq@gmail.com}
\affiliation{Tencent Quantum Laboratory, Tencent, Shenzhen, Guangdong 518057, China}

\author{Shengyu Zhang}
\email{shengyzhang@tencent.com}
\affiliation{Tencent Quantum Laboratory, Tencent, Shenzhen, Guangdong 518057, China}

\begin{abstract}
High-fidelity readout with strong Purcell protection of qubit coherence is essential for scalable superconducting quantum processors, yet the finite passband and sizable footprint of conventional band-pass Purcell filters make them hard to scale. Here we introduce a scalable edge-pass Purcell filter that separates the readout band from the protected qubit band by a single transmission edge, freeing the readout resonators from bandwidth constraint. Depending on whether the transmitting band lies above or below the cutoff, the compact network is realized as a high-pass filter (HPF) or a low-pass filter (LPF). The HPF reaches an average readout fidelity of $99.46(4)\%$ (up to $99.56\%$) with a $150$-ns pulse, and the LPF reaches $99.49(3)\%$ (up to $99.57\%$) with a $130$-ns pulse. The average single-qubit gate fidelities are $99.94\%$ (HPF) and $99.93\%$ (LPF). Relative to the filter-free Purcell limit, the filters substantially extend the qubit lifetime, and the Purcell protection deepens at higher filter order. In addition, an intrinsic dissipation mode of the filter offers a qubit-reset channel. This leads to a compact architecture that unifies fast, high-fidelity readout, Purcell protection, and effective reset within a single filter for large-scale fault-tolerant quantum computation.
\end{abstract}

\maketitle

\section{Introduction}\label{Sec1}

Quantum error correction\cite{fowler2012surface, google2023suppressing, acharya2025quantum} in large-scale superconducting quantum processors\cite{arute2019quantum, bravyi2022future} requires the qubits to be read out quickly and accurately in every cycle. At the same time, the readout should not degrade the coherence of the measured qubit. In the widely used dispersive readout scheme\cite{blais2004cavity, koch2007charge}, the qubit state is encoded in a state-dependent frequency shift of a readout resonator, and a strong qubit--resonator coupling\cite{wallraff2004strong} is desirable for fast and accurate state discrimination. However, this coupling also opens a Purcell channel\cite{koch2007charge, houck2008controlling} through which the qubit excitation decays into the readout line, so that a stronger coupling which speeds up the readout will shorten the qubit lifetime. Therefore, achieving fast and high-fidelity readout\cite{johnson2012heralded, jeffrey2014fast, walter2017rapid, heinsoo2018rapid, sunada2022fast, chen2023transmon, spring2024fast, sunada2024photon, wang202499, chapple2025balanced, xiong2025high, beaulieu2026fast} while preserving a long qubit coherence remains a central challenge for scalable superconducting quantum computation.

A common way to reconcile these conflicting requirements is to insert a Purcell filter\cite{sete2015quantum} between the readout resonator and the transmission line. The filter shapes the electromagnetic environment seen by the qubit, suppressing the radiative loss at the qubit frequency while leaving the resonator strongly coupled to the line within the readout band, and thus allows fast readout without a severe damage in qubit lifetime. However, most Purcell filters demonstrated so far are belonging to band-pass structures, realized as single-resonator filters\cite{jeffrey2014fast, chen2023transmon}, notch filters\cite{reed2010fast, gu2025engineering}, coupled-resonator filters\cite{mohebbi2014superconducting, bronn2015broadband, naaman2022synthesis, iakoupov2023saturable, yan2023broadband, zhou2024high,  park2024characterization, thorbeck2024high, luo2026compact} or other designs\cite{yen2024all, bakr2024multiplexed, yen2025interferometric, xiao2025flexible, ahmad20263d}. These designs work well, but their finite bandwidth and the sizable footprint of the coplanar-waveguide (CPW) resonators limit how many readout resonators fit into the band and complicate the frequency allocation and wiring as the qubit number grows. One route explored to ease this limitation is a diplexer\cite{ding2024multi} that combines a high-pass\cite{smitham2025sub} and a low-pass filter\cite{chen2023superconducting}, using the high-pass branch for readout and the low-pass dissipator for qubit reset. This scheme relaxes the bandwidth limitation to some extent, but it comes at the cost of a bulky composite structure that complicates the qubit interconnection, so that the trade-off between bandwidth and scalability remains only partly resolved.

\begin{figure*}[htb]
\includegraphics{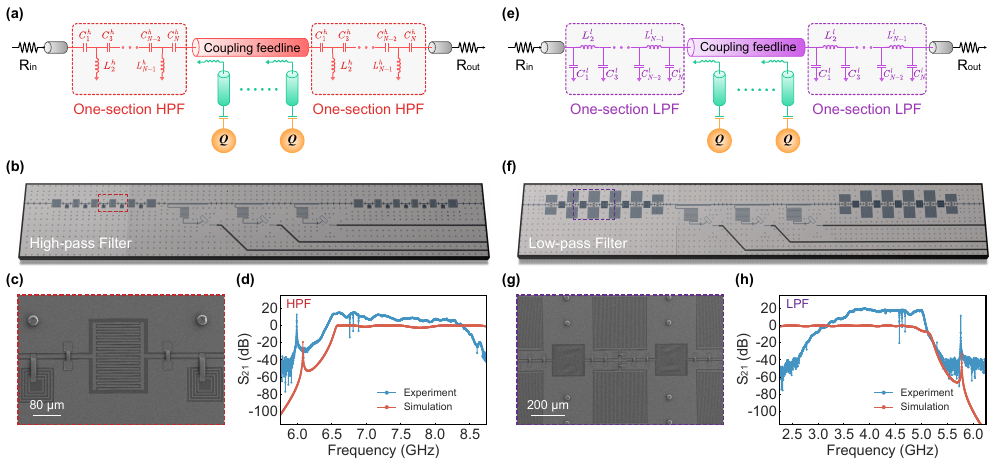}
\caption{\textbf{Overview of the edge-pass Purcell filters.} (a)(e) Circuit schematics of the high-pass filter (HPF) and low-pass filter (LPF). The red (purple) dashed box denotes the circuit structure of the one-section HPF (LPF), and the coupling feedline couples to the readout resonator (green) and the qubit (orange). (b)(f) Optical micrographs of the $11$-order two-section HPF and LPF in the flip-chip devices, both residing on the carrier chip (the qubit-layer top chip is not shown). (c)(g) Magnified SEM images of the interdigital capacitors and spiral inductors used in the HPF and LPF. (d)(h) Measured (blue) and finite-element-simulated (red) transmission spectra $S_{21}$ of the $11$-order two-section HPF and LPF.}
\label{fig:Fig1}
\end{figure*}

In this work, we address this trade-off with a scalable edge-pass Purcell filter that uses the transmission edge of the filter, rather than a bounded passband, to separate the two bands. The readout band and the protected qubit band sit on opposite sides of a single cutoff, so that the readout resonators are no longer confined to a limited bandwidth and can be allocated freely. Depending on whether the transmitting band lies above or below the cutoff, the filter takes the form of a high-pass filter (HPF) or a low-pass filter (LPF). Each is built from a compact two-section lumped-element network that keeps the footprint small and the layout regular. In experiment, the HPF reaches an average single-shot\cite{mallet2009single} readout fidelity of $99.46(4)\%$ (up to $99.56\%$) with a $150$-ns readout pulse, and the LPF reaches $99.49(3)\%$ (up to $99.57\%$) with a $130$-ns pulse. The average single-qubit gate fidelities are $99.94\%$ (HPF) and $99.93\%$ (LPF). These results show that the edge-pass design is fully compatible with high-performance qubit operation. By comparing the measured relaxation times with the Purcell-limited values, we confirm that the filter provides effective Purcell protection, which strengthens as the filter order increases. We further find that the same filter naturally hosts an intrinsic dissipation mode that can be used to reset the qubit, and, building on this, we propose a scalable architecture in which high-performance readout, strong Purcell protection, and effective qubit reset are integrated within a single filter.

\section{Design and characterizations}\label{Sec2}

We begin by describing the circuit architecture of the edge-pass filters and characterizing their transmission properties. The guiding idea is to separate the readout band from the protected qubit band by a single transmission edge rather than by a bounded passband, so that the two bands sit on opposite sides of one cutoff frequency and the readout resonators can be distributed freely on the transmitting side. Both filters share a common building block: a compact two-section network in which identical one-section units terminate the input and output ports and a coupling feedline bridges them, as depicted in Figs.\ \ref{fig:Fig1}(a) and (e). Here a section refers to one such lumped-element unit placed at a port; the two-section network thus carries one section at each end, whereas a one-section filter carries a single unit at one port only (see Supplementary Materials). Each one-section unit is a Chebyshev type-I filter\cite{pozar2012microwave}, because it provides a sharp roll-off at the cutoff and a stopband whose rejection deepens monotonically with the filter order; the two-section arrangement then places two such sharp edges on either side of the feedline, which sharpens the overall transmission edge and deepens the rejection in the protected band. A device is labelled by the order $N$ of these one-section units as an $N$-order HPF or LPF (unless stated otherwise, HPF and LPF denote the full two-section filters throughout this work). The central feedline couples to the readout resonator, and thereby to the qubit, forming the measurement channel. Terminating the feedline with two filter sections imposes boundary conditions that give rise to an intrinsic standing-wave mode [Figs.\ \ref{fig:Fig1}(d) and (h)], hereafter referred to as the ``dissipation mode"\cite{ding2024multi}, which is later used as a reset channel. This single building block covers both configurations: placing the transmitting band above or below the cutoff yields the HPF or the LPF, so that the same design procedure applies to both. To characterize the design across a range of orders, three chips were fabricated and measured (see the Methods for the detailed fabrication procedure): a planar HPF (P-HPF) hosting $5$ to $11$-order filters, and flip-chip HPF (F-HPF) and LPF (F-LPF) chips hosting $11$ to $13$-order filters.

For the HPF, the cutoff frequency was set to $\omega_c^h/2\pi\approx6.5$~GHz, a value that leaves room for the readout resonators in the transmitting band above it while keeping the qubit band well inside the stopband below it. This choice fixes the section capacitances $C_k^h$ and inductances $L_k^h$ through\cite{pozar2012microwave}
\begin{equation}
C_k^h=\frac{1}{Z_0\omega_c^h g_k},~~L_k^h=\frac{Z_0}{\omega_c^h g_k}~~(k=1\cdots N),
\tag{1}
\label{eq.1}
\end{equation}
where $g_k$ are the low-pass prototype values (see Methods) and $Z_0$ is the characteristic impedance. These lumped elements were realized using interdigital capacitors and spiral inductors, as shown in Figs.\ \ref{fig:Fig1}(b) and (c). Because the response is set by the lumped $L$ and $C$ values rather than by a resonator whose size scales with the wavelength, the filter is substantially more compact than a CPW-resonator filter of comparable performance, which is essential for integrating many filters on a single chip. The readout resonators were placed at $6.7$--$6.9$~GHz, above the cutoff frequency and hence within the transmitting band, whereas the maximum qubit frequency was set near $4.9$~GHz within the protected stopband. The feedline length was fixed at about $5$~mm, long enough to accommodate multiple readout resonators while keeping the dissipation mode far detuned from the qubit frequency to minimize its contribution to qubit decay (see Supplementary Materials). As illustrated in Fig.\ \ref{fig:Fig1}(d), we used finite-element method to simulate and then measured the transmission spectrum of the $11$-order HPF, and the two agree well, confirming that the lumped-element design behaves as intended.

The LPF follows by duality: the transmitting and protected bands are exchanged, so the readout resonators now sit below the cutoff and the qubit band above it. The cutoff frequency was set to $\omega_c^l/2\pi=5$~GHz, giving\cite{pozar2012microwave}
\begin{equation}
C_k^l=\frac{Z_0 g_k}{\omega_c^l},~~L_k^l=\frac{g_k}{Z_0\omega_c^l}~~(k=1\cdots N).
\tag{2}
\label{eq.2}
\end{equation}
Compared with the HPF, this configuration calls for larger capacitances and inductances. To accommodate these larger element values, each capacitor was implemented as a pair of parallel interdigital capacitors, and each inductor as a double spiral architecture, as shown in Figs.\ \ref{fig:Fig1}(f) and (g), which reaches the required values without resorting to excessive airbridges and thereby reduces the insertion loss. The frequency of readout resonators was placed at $4.5$--$4.7$~GHz, below the cutoff frequency, and the maximum qubit frequency was set near $6.3$~GHz. Following the same consideration as for the HPF, the coupling feedline length was also set to about $5$~mm. Likewise, we simulated the transmission spectrum of the $11$-order LPF using finite-element method and then measured it [Fig.\ \ref{fig:Fig1}(h)]. As seen in Figs.\ \ref{fig:Fig1}(d) and (h), the transmission at the qubit frequency is strongly suppressed in both filters while remaining high across the readout band, so that the readout is left unaffected while an effective Purcell protection of the qubit can be anticipated.

\begin{figure}[htb]
\includegraphics[width=85mm, height=124mm]{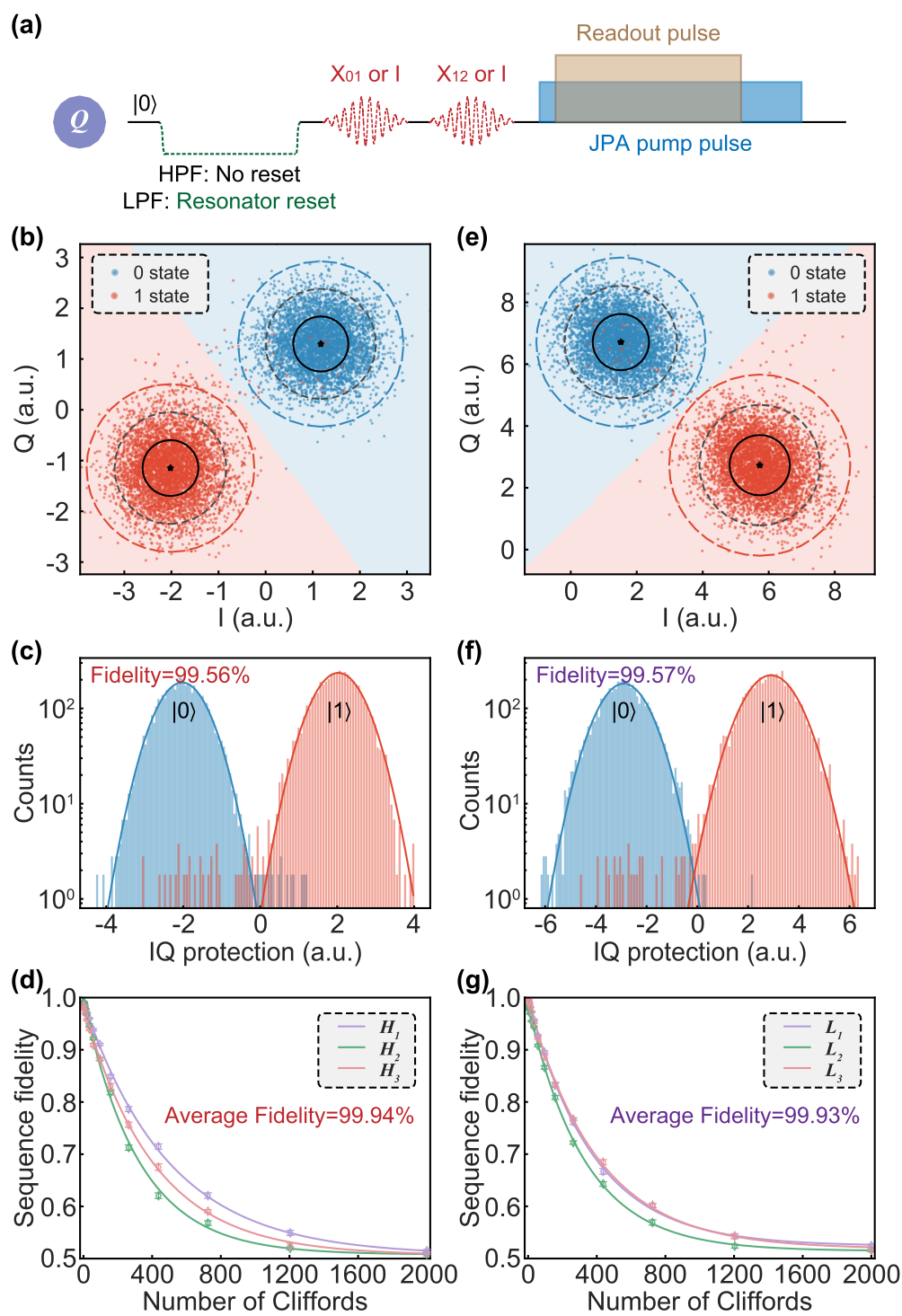}
\caption{\textbf{Readout and single-qubit gate characterization.} (a) Readout sequence. The qubit is prepared with $X_{01}$ and $X_{12}$ pulses (or identity operations $I$) for multilevel readout, followed by the readout pulse together with the JPA\cite{roy2015broadband, sun2025broadband} pump pulse; no pre-readout reset is applied for the HPF, whereas a resonator-based reset is applied for the LPF. (b)(e) Single-shot scatter plots of the readout signal in the IQ plane for the HPF and LPF, showing well-separated $|0\rangle$ and $|1\rangle$ distributions. (c)(f) Histograms of the IQ projection, from which the readout fidelities of $99.56\%$ (HPF) and $99.57\%$ (LPF) are obtained. (d)(g) Randomized benchmarking of the single-qubit gates for three qubits in the HPF and LPF, giving average gate fidelities of $99.94\%$ and $99.93\%$, respectively.}
\label{fig:Fig2}
\end{figure}

\section{Experimental demonstrations}\label{Sec3}

With the circuit design and transmission characteristics of the edge-pass filters described above, we now study how qubits perform when coupled to them. Qubits and readout resonators were integrated with both the HPF and the LPF, and we characterized two aspects that are essential for a practical readout architecture: the quality of the readout and single-qubit control, and the Purcell protection of the qubit. We first benchmark the readout and single-qubit gates in both filters, and then quantify the Purcell protection that the filters provide to the qubit lifetime.

\subsection{High-fidelity readout and single-qubit control}\label{Sec3.1}

For a practical Purcell filter, it should protect the qubit without compromising the readout, so we first characterized the qubit readout in both the HPF and the LPF. Because the readout resonators lie within the transmitting band, the readout signal passes through the filter with little attenuation, and the edge-pass filters therefore can support fast and high-fidelity single-shot dispersive readout\cite{mallet2009single, walter2017rapid}. The readout sequences are illustrated in Fig.\ \ref{fig:Fig2}(a), where a multilevel readout protocol\cite{chen2023transmon, xiong2025high} was employed in each case, i.e., the qubit state was discriminated with the higher transmon levels included in the classification to suppress the assignment error caused by measurement-induced leakage\cite{sank2016measurement}. In the HPF, the qubit state was measured directly with a $150$-ns readout pulse, whereas in the LPF a short pre-readout exchange with the resonator mode initialized the qubit before a $130$-ns readout pulse was applied. The resulting single-shot scatter plots in the IQ plane and the corresponding projection histograms are shown in Figs.\ \ref{fig:Fig2}(b)(e) and Figs.\ \ref{fig:Fig2}(c)(f), respectively; in both filters the distributions of the different states are well separated, allowing the qubit state to be assigned from a single measurement. The readout fidelity is quantified as $F=\frac{1}{2}\left[P(0|g)+P(1|e)\right]$, where $P(0|g)$ ($P(1|e)$) is the probability of obtaining outcome $|0\rangle$ ($|1\rangle$) for a qubit prepared in $|g\rangle$ ($|e\rangle$). Averaged over the qubits of the $11$-order devices, the F-HPF reached a readout fidelity of $99.46(4)\%$, with the best qubit at $99.56\%$. The F-LPF reached $99.49(3)\%$ on average, with the best qubit at $99.57\%$. These results confirm that the edge-pass filter supports high-fidelity readout in both the high-pass and low-pass configurations.

We next characterized the single-qubit gates in both filters. Using randomized benchmarking (RB)\cite{magesan2012characterizing, magesan2012efficient}, in which sequences of random Clifford gates of increasing length are applied and the surviving fidelity is measured, we benchmarked three qubits in each device [Figs.\ \ref{fig:Fig2}(d)(g)]. The extracted average single-qubit gate fidelities were $99.94\%$ (F-HPF) and $99.93\%$ (F-LPF), reaching a high level in both filters. Together with the readout results above, these results demonstrate that the edge-pass filters are fully compatible with high-fidelity single-qubit operation.

\begin{figure}[htb]
\includegraphics{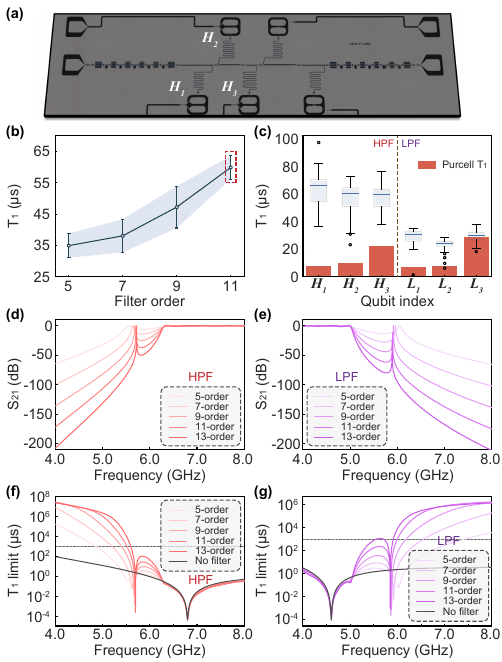}
\caption{\textbf{Purcell protection of the edge-pass filters.} (a) Optical micrograph of the $11$-order two-section HPF on the planar chip, with three tunable qubits $H_1\sim H_3$ and two fixed-frequency qubits. (b) Measured $T_1$ versus filter order for the $5$- to $11$-order HPFs; each point is the mean over the five qubits of a group. (c) Measured $T_1$ compared with the Purcell $T_1$ limit for the HPF and LPF. Left of the dashed line: the $11$-order planar HPF ($H_1\sim H_3$); right: the $11$-order flip-chip LPF ($L_1\sim L_3$). (d)(e) Simulated $S_{21}$ of the $5$- to $13$-order HPFs and LPFs. (f)(g) Corresponding simulated Purcell-limited relaxation times; the horizontal dashed line marks $T_1=1$~ms.}
\label{fig:Fig3}
\end{figure}

\subsection{Purcell protection of qubit coherence}\label{Sec3.2}

The high readout and gate fidelities above confirm that the qubits operate well within the edge-pass filters. Next, we turn to the Purcell protection of the qubit. When a qubit is dispersively coupled to a readout resonator, it acquires a radiative decay channel into the transmission line\cite{houck2008controlling, sete2015quantum}, and the filter is designed to suppress this loss by presenting a high impedance to the environment at the qubit frequency. The strong suppression of $S_{21}$ at the qubit frequency shown in Fig.\ \ref{fig:Fig1} therefore indicates that the filters can shield the qubit from Purcell decay, which we confirmed experimentally. Since the stopband becomes steeper with increasing filter order [Figs.\ \ref{fig:Fig3}(d)(e)], the suppression at the qubit frequency deepens and the protection is expected to strengthen with the order $N$. To verify this expectation, we performed measurements on the P-HPF chip [Fig.\ \ref{fig:Fig3}(a)], which carries $5$ to $11$-order filters sharing identical qubits and resonators distributions, so that the filter order is the only variable across the groups. As shown in Fig.\ \ref{fig:Fig3}(b), the measured average qubits lifetimes increased systematically with the filter order, in agreement with the expectation.

To quantify how strong the Purcell protection is, we compared the measured lifetimes with the Purcell limit\cite{sete2015quantum} expected in the absence of a filter:
\begin{equation}
T_1^{\mathrm{no~filter}}=\frac{\Delta_{qr}^2}{\kappa_r g_{qr}^2},
\tag{3}
\label{eq.3}
\end{equation}
where $\Delta_{qr}=\omega_q-\omega_r$ is the qubit--resonator detuning, $\kappa_r$ is the resonator linewidth, and $g_{qr}$ is the qubit--resonator coupling strength. This expression sets the relaxation time that the same qubit and resonator would exhibit if the resonator were directly coupled to the environment, and thus serves as a reference against which the benefit of the filter can be assessed. Figure \ref{fig:Fig3}(c) compares the measured lifetimes of the $11$-order HPF ($H_1\sim H_3$) and LPF ($L_1\sim L_3$) with this unfiltered limit. The corresponding resonator linewidths are $\kappa_r/2\pi=6.85$, $5.58$, and $2.58$~MHz for $H_1\sim H_3$ and $\kappa_r/2\pi=10.00$, $12.61$, and $3.00$~MHz for $L_1\sim L_3$, which enter Eq.\ \eqref{eq.3} to set the filter-free Purcell limit of each qubit. In both filters, the measured relaxation times clearly exceed the corresponding filter-free Purcell limit, indicating that the edge-pass filters substantially suppress the qubit relaxation through the readout channel and thereby provide an effective Purcell protection.

To obtain a quantitative and predictive description of the protection, we modelled the readout circuits [Fig.\ \ref{fig:Fig1}(a)] using the black-box quantization method\cite{nigg2012black}, in which the entire linear network seen by the qubit is represented by its admittance $Y_q(\omega)$ at the qubit port. The Purcell-limited lifetime then follows as\cite{nigg2012black}
\begin{equation}
T_1^{\mathrm{Purcell}}=\frac{C_q}{\mathrm{Re}[Y_q(\omega_q)]},
\tag{4}
\label{eq.4}
\end{equation}
where $C_q$ is the effective qubit capacitance and $Y_q$ is the admittance seen from the qubit port, so that a smaller dissipative admittance at $\omega_q$ corresponds to a longer Purcell-limited lifetime. The simulated transmission spectra of the HPF ($\omega_c^h/2\pi\approx6.3$~GHz) and LPF ($\omega_c^l/2\pi\approx5.0$~GHz) are shown in Figs.\ \ref{fig:Fig3}(d) and (e), and the corresponding $T_1$ limits, evaluated with readout resonators at $6.8$~GHz (HPF) and $4.6$~GHz (LPF), each with external quality factor $Q_c\sim 500$, in Figs.\ \ref{fig:Fig3}(f) and (g). Each case exhibits two dips in the $T_1$ limit, one at the readout resonator ($6.8$ and $4.6$~GHz for the HPF and LPF, respectively), where the qubit is directionally coupled to the readout resonator, and the other at the dissipation mode ($\sim5.7$~GHz). Away from these modes, the simulated $T_1$ limit flattens, at low frequency for the HPF and at high frequency for the LPF, reflecting a floor set by the substrate loss assumed in the model rather than by the Purcell effect; the Purcell protection itself would otherwise continue to improve away from the passband. The predicted lifetimes confirmed the measured trends and their improvement with the order, in agreement with the experimental results. Finally, the pronounced $T_1$ dip at the dissipation mode identifies a natural dissipation channel intrinsic to the filter, which we exploit for the qubit reset below.

\begin{figure}[htb]
\includegraphics{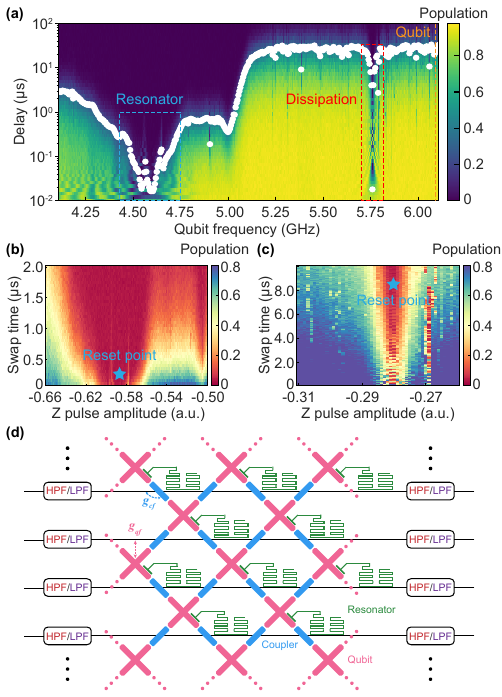}
\caption{\textbf{Qubit reset and envisioned scalable architecture.} (a) Qubit population as a function of the qubit frequency and the delay time, from which the qubit relaxation time is extracted (white dots). Two dips appear, marked by the blue and red dashed boxes, corresponding to the resonator mode and the dissipation mode, respectively; the orange dashed line marks the qubit sweet spot. (b)(c) Residual qubit population as a function of the $Z$ pulse amplitude and the swap time, for reset via the resonator mode and the dissipation mode, respectively. The reset point, at which the population is minimized, is marked in each panel. (d) Envisioned scalable architecture unifying fast and high-performance readout, Purcell protection, and effective reset, where the qubit couples to the edge-pass filter (HPF or LPF) either directly with strength $g_{qf}$ or through its coupler with strength $g_{cf}$.}
\label{fig:Fig4}
\end{figure}

\section{Discussions}\label{Sec4}

Besides the high-fidelity readout and Purcell protection, our edge-pass filters provide an intrinsic dissipation mode that can be used as a built-in channel for qubit reset\cite{ding2024multi, gu2025engineering} without any additional hardware. As identified in the Purcell-protection analysis above, the filter supports two distinct dissipative modes that a qubit can be brought into resonance with: the readout-resonator mode and the dissipation mode. Taking the $11$-order LPF as an example, we can see from Fig.\ \ref{fig:Fig4}(a) that the measured relaxation times exhibit a dip at both modes, in agreement with the simulation in Fig.\ \ref{fig:Fig3}(g), which shows that both modes can be used to reset the qubit. To reset the qubit, we tuned its frequency into resonance with both modes respectively by applying a $Z$ pulse, so that the qubit excitation was swapped into the mode and dissipated. Figure \ref{fig:Fig4}(b) and (c) show the residual qubit population as a function of the $Z$ pulse amplitude and the swap time, for the resonator mode and the dissipation mode, respectively; the reset point, at which the population is minimized, is marked in each panel. Using the resonator mode, the qubit was reset with a swap time of about $150$~ns, whereas the dissipation mode required a longer swap time of about $8.5~\mu$s to reach a comparable residual population. The difference in timescale reflects the linewidths of the two modes: the resonator mode is strongly coupled to the transmission line and therefore dissipates the excitation rapidly, while the dissipation mode is comparatively narrow. Importantly, the linewidth of the dissipation mode is set by the filter order and can be broadened by decreasing it, so that the dissipation-mode reset can in principle be accelerated\cite{ding2024multi}. The reset demonstrated here is therefore not a separately engineered function but a capability already inherent to the filter that performs readout and Purcell protection.

The ability of an edge-pass filter to support fast and high-performance readout, Purcell protection, and effective reset simultaneously suggests a natural route toward scaling. As illustrated in Fig.\ \ref{fig:Fig4}(d), a qubit can be coupled to the dissipation mode of the same edge-pass filter, either an HPF or an LPF, in two ways: directly with a coupling strength $g_{qf}$\cite{ding2024multi, gu2025engineering}, or indirectly through a tunable coupler with a coupling strength $g_{cf}$\cite{xiao2025flexible}. In either case, the purpose is to open a dissipative channel from the qubit to the filter so that its excitation can be drained rapidly, thereby speeding up the reset. The direct coupling provides a strong and simple channel for fast reset, whereas the coupler-mediated scheme introduces the dissipation indirectly, which accelerates the reset while minimizing its impact on the qubit relaxation. Within both schemes, the transmitting band serves fast and high-fidelity readout, the stopband provides Purcell protection, and the dissipation mode enables qubit reset. By integrating these three functions into a single compact filter, this architecture reduces both the footprint and the wiring overhead per qubit, which are among the main obstacles to scaling superconducting processors. This makes the edge-pass filter a promising hardware primitive for the repeated measurement and reset operations required by quantum error correction\cite{mcewen2021removing, krinner2022realizing}, and its planar and flip-chip realizations indicate compatibility with the integration schemes used in large-scale devices.

In summary, we have proposed and demonstrated scalable edge-pass Purcell filters that separate the readout band from the protected qubit band by a single transmission edge, so that the readout resonators are free from any bandwidth constraint while the filter keeps a compact footprint. The two-section building block yields both a high-pass filter (HPF) and a low-pass filter (LPF), and in both we measured fast and high-performance readout. The HPF reached an average readout fidelity of $99.46(4)\%$ (up to $99.56\%$) in $150$~ns, and the LPF reached $99.49(3)\%$ (up to $99.57\%$) in $130$~ns. Together with average single-qubit gate fidelities of $99.94\%$ (HPF) and $99.93\%$ (LPF), these results show that the filters are fully compatible with high-performance qubit operation. Comparing the measured relaxation times with the filter-free Purcell limit, we found that the filters substantially extend the qubit lifetime, and the protection deepens as the filter order increases. We further identified an intrinsic dissipation mode of the same filter and used it to reset the qubit, and on this basis proposed an architecture in which fast and high-fidelity readout, Purcell protection, and effective reset are integrated within a single filter. The same design also extends to asymmetric configurations, which use sections of different order at the two ports to add flexibility (the details are given in the Supplementary Materials). While these results establish the edge-pass filter as a compact and versatile element, accelerating the dissipation-mode reset and experimentally validating the fully integrated architecture remain to be addressed. Together with extending the design to a full multi-qubit module, this would make the edge-pass filter a practical building block for large-scale fault-tolerant superconducting quantum processors.

\section{Methods}

\subsection{Design parameters}

\begin{table*}[htb]
\centering
\caption{Low-pass filter prototype values $g_k$ used in Eqs.\ \eqref{eq.1} and \eqref{eq.2} for different filter orders.}
\label{tab:1}
\renewcommand{\arraystretch}{1.4}
\begin{tabular*}{\textwidth}{@{\extracolsep{\fill}} c c c c c c c c c c c c c c}
\hline\hline
Order & $g_1$ & $g_2$ & $g_3$ & $g_4$ & $g_5$ & $g_6$ & $g_7$ & $g_8$ & $g_9$ & $g_{10}$ & $g_{11}$ & $g_{12}$ & $g_{13}$ \\
\hline
5  & 1.7058 & 1.2296 & 2.5408 & 1.2296 & 1.7058 &        &        &        &        &        &        &        &        \\
7  & 1.7372 & 1.2583 & 2.6381 & 1.3444 & 2.6381 & 1.2583 & 1.7372 &        &        &        &        &        &        \\
9  & 1.7504 & 1.2690 & 2.6678 & 1.3673 & 2.7239 & 1.3673 & 2.6678 & 1.2690 & 1.7504 &        &        &        &        \\
11 & 1.7572 & 1.2743 & 2.6809 & 1.3759 & 2.7488 & 1.3879 & 2.7488 & 1.3759 & 2.6809 & 1.2743 & 1.7572 &        &        \\
13 & 1.7610 & 1.2772 & 2.6878 & 1.3802 & 2.7596 & 1.3955 & 2.7714 & 1.3955 & 2.7596 & 1.3802 & 2.6878 & 1.2772 & 1.7610 \\
\hline\hline
\end{tabular*}
\end{table*}

In the filter-synthesis approach, a filter is first described by a normalized low-pass prototype, a ladder of unit-less elements $g_k$ set by the filter order $N$ and the chosen prototype family. Several prototype families are commonly used, including the Butterworth, Chebyshev type-I, Chebyshev type-II, elliptic, and Bessel filters\cite{wyndrum1965microwave, pozar2012microwave}. They differ mainly in the behavior of the passband and in the steepness of the transition band, which is also tied to the phase response of the filter. Balancing these considerations, we adopt the Chebyshev type-I prototype with a $0.5$-dB passband ripple, which offers a sufficiently sharp transition at the cutoff while keeping the in-band ripple and the phase distortion moderate. To obtain a physical filter, the prototype is scaled to the desired cutoff frequency and to the port impedance. This scaling converts the dimensionless $g_k$ into the capacitances and inductances of Eqs.\ \eqref{eq.1} and \eqref{eq.2}. Table \ref{tab:1} lists the prototype coefficients $g_k$ for the orders $N=5,7,9,11,13$ used in this work. Given these coefficients, together with the port impedance and the chosen cutoff, the lumped-element values of each HPF and LPF are fully determined.

\subsection{Fabrication procedure}

The devices are patterned from a $200$-nm-thick $\alpha$-tantalum film sputtered on a sapphire substrate. The film is defined by photolithography and etching, which forms the coplanar waveguides, the readout resonators, the qubit capacitors, and the lumped elements of the edge-pass filters. The Josephson junctions with the Manhattan style are then formed by double-angle evaporation. Airbridges\cite{bu2025tantalum} are fabricated by lift-off method, they suppress parasitic slotline modes and stray coupling across the control and readout lines. In the HPF, they also connect the inner terminals of the spiral inductors.

The planar device integrates the filters, resonators, and qubits on a single chip. For the flip-chip devices, the qubits are placed on a top chip while the readout resonators, control lines, and edge-pass filters are fabricated on a separate carrier chip. Dense indium bumps of $9~\mu$m in height are deposited on both chips. The two chips are then aligned and flip bonded together\cite{liao2026breaking}.

\section{Data Availability}
The data that support the findings of this study are available from the corresponding authors X.P.Y., T.Q.C. and S.Y.Z. upon request.

\section{Code Availability}
The code that supports the simulations of this study are available from the corresponding authors X.P.Y., T.Q.C. and S.Y.Z. upon request.

\section{Acknowledgements} 
We thank Fuming Liu, Guanglei Xi, Qiaonian Yu and Hualiang Zhang for supporting room-temperature electronics. 

\section{Author Contributions}
T.Q.C. and X.D.L. proposed the idea and conceived the experiment. X.D.L., T.Q.C., and X.P.Y. designed the device. Y.L. fabricated the devices with assistance of T.Q.C., X.P.Y., Z.X.Z., S.N.H. and K.L.B.. T.Q.C. established the measurement setup and performed experimental measurements with assistance of X.D.L., X.P.Y., S.N.H., Z.X.Z., S.Y.P., Z.W.Z., Y.L.Y. and S.Y.Z.. X.D.L. performed the theoretical analyses and numerical simulations with assistance of T.Q.C. and S.N.H.. X.D.L wrote the manuscript with feedback from all authors. X.P.Y., T.Q.C. and S.Y.Z. supervised the project. All authors contributed to the discussion of the results and development of the manuscript.

\makeatletter
\renewcommand\@biblabel[1]{[#1]}
\makeatother

%

\bigskip

\end{document}


\title{Supplementary Materials: A scalable edge-pass Purcell filter for high-fidelity readout of superconducting qubits}

\author{Xudong Liao}
\altaffiliation{These three authors contributed equally to this work.}
\affiliation{National Laboratory of Solid State Microstructures, School of Physics, Nanjing University, Nanjing, 210093 Jiangsu, China}
\affiliation{Shishan Laboratory, Nanjing University, Suzhou, 215163 Jiangsu, China}
\affiliation{Jiangsu Key Laboratory of Quantum Information Science and Technology, Nanjing University, Suzhou, 215163 Jiangsu, China}

\author{Yuan Li}
\altaffiliation{These three authors contributed equally to this work.}
\affiliation{Tencent Quantum Laboratory, Tencent, Shenzhen, Guangdong 518057, China}

\author{Sainan Huai}
\altaffiliation{These three authors contributed equally to this work.}
\affiliation{Tencent Quantum Laboratory, Tencent, Shenzhen, Guangdong 518057, China}

\author{Shuyi Pan}
\affiliation{National Laboratory of Solid State Microstructures, School of Physics, Nanjing University, Nanjing, 210093 Jiangsu, China}
\affiliation{Shishan Laboratory, Nanjing University, Suzhou, 215163 Jiangsu, China}
\affiliation{Jiangsu Key Laboratory of Quantum Information Science and Technology, Nanjing University, Suzhou, 215163 Jiangsu, China}

\author{Zhenxing Zhang}
\affiliation{Tencent Quantum Laboratory, Tencent, Shenzhen, Guangdong 518057, China}

\author{Zhiwen Zong}
\affiliation{Tencent Quantum Laboratory, Tencent, Shenzhen, Guangdong 518057, China}

\author{Kunliang Bu}
\affiliation{Tencent Quantum Laboratory, Tencent, Shenzhen, Guangdong 518057, China}

\author{Yulei Ye}
\affiliation{Tencent Quantum Laboratory, Tencent, Shenzhen, Guangdong 518057, China}

\author{Wen Zheng}
\affiliation{National Laboratory of Solid State Microstructures, School of Physics, Nanjing University, Nanjing, 210093 Jiangsu, China}
\affiliation{Shishan Laboratory, Nanjing University, Suzhou, 215163 Jiangsu, China}
\affiliation{Jiangsu Key Laboratory of Quantum Information Science and Technology, Nanjing University, Suzhou, 215163 Jiangsu, China}

\author{Xinsheng Tan}
\affiliation{National Laboratory of Solid State Microstructures, School of Physics, Nanjing University, Nanjing, 210093 Jiangsu, China}
\affiliation{Shishan Laboratory, Nanjing University, Suzhou, 215163 Jiangsu, China}
\affiliation{Jiangsu Key Laboratory of Quantum Information Science and Technology, Nanjing University, Suzhou, 215163 Jiangsu, China}

\author{Yang Yu}
\affiliation{National Laboratory of Solid State Microstructures, School of Physics, Nanjing University, Nanjing, 210093 Jiangsu, China}
\affiliation{Shishan Laboratory, Nanjing University, Suzhou, 215163 Jiangsu, China}
\affiliation{Jiangsu Key Laboratory of Quantum Information Science and Technology, Nanjing University, Suzhou, 215163 Jiangsu, China}

\author{Xiaopei Yang}
\email{xiaopeiyang@tencent.com}
\affiliation{Tencent Quantum Laboratory, Tencent, Shenzhen, Guangdong 518057, China}

\author{Tianqi Cai}
\email{tsefctq@gmail.com}
\affiliation{Tencent Quantum Laboratory, Tencent, Shenzhen, Guangdong 518057, China}

\author{Shengyu Zhang}
\email{shengyzhang@tencent.com}
\affiliation{Tencent Quantum Laboratory, Tencent, Shenzhen, Guangdong 518057, China}

\maketitle

\tableofcontents

\newpage

\section{Experimental setup}\label{SecS1}

\begin{figure}[htb]
\includegraphics{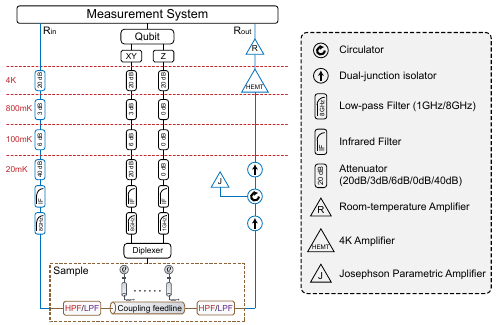}
\caption{\textbf{Schematic of the measurement setup.} The input, $XY$, and $Z$ lines are attenuated and filtered across the temperature stages of the dilution refrigerator, and the readout output is amplified by a Josephson parametric amplifier, a high-electron-mobility transistor amplifier, and a room-temperature amplifier. The sample sits at the base stage.}
\label{fig:FigS1}
\end{figure}

The measurement setup\cite{cai2025multiplexed} is shown in Fig.\ \ref{fig:FigS1}. The chip was mounted in an aluminum sample box anchored to the base-temperature stage of a dilution refrigerator, with infrared and magnetic shielding around it. The control and readout waveforms were generated by room-temperature electronics\cite{zhang2021exploiting}. The qubit $XY$ drives were produced by IQ modulation, and the $Z$-control pulses were synthesized directly. The probe, $XY$, and $Z$ input lines were attenuated and filtered at the successive temperature stages to reduce thermal noise and high-frequency interference. On the output side, the readout signal first passed through circulators and isolators, and was then amplified by a Josephson parametric amplifier (JPA)\cite{roy2015broadband, sun2025broadband} at the base stage, a high-electron mobility transistor amplifier at the $4$~K stage, and a room-temperature amplifier. The amplified signal was finally demodulated and digitized.

\section{Filter simulation}\label{SecS2}

\subsection{Characterization of the dissipation mode}\label{SecS2.1}

\begin{figure}[tb]
\includegraphics{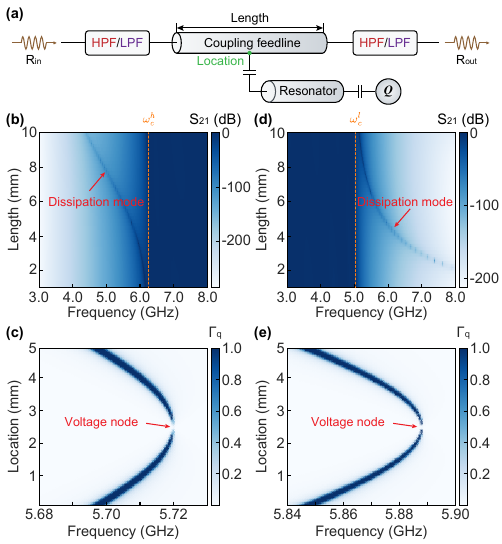}
\caption{\textbf{Simulation of the dissipation mode.} (a) Schematic of the simulation circuit. ``Length'' is the coupling-feedline length, and ``Location'' is the position of the resonator and qubit along the feedline. (b)(d) Simulated $S_{21}$ of the $13$-order HPF and LPF, without the resonator and qubit, as a function of feedline length; the red arrows indicate the dissipation mode and the orange dashed lines the cutoff frequencies $\omega_c^h$ and $\omega_c^l$. (c)(e) Normalized qubit dissipation $\Gamma_q$ versus the resonator-qubit location for the $13$-order HPF and LPF at a $5$~mm feedline length; the red arrows mark the voltage node, at which the qubit decouples from the dissipation mode.}
\label{fig:FigS2}
\end{figure}

Terminating the coupling feedline with two filter sections creates a standing-wave mode on the feedline, which we refer to as the dissipation mode. This mode is lossy because both of its ends see the external environment through the filters, and it plays a dual role depending on its frequency. If the dissipation mode overlaps the qubit band, it opens an extra relaxation channel and degrades coherence, so it must be kept clear of that band. If the dissipation mode is instead tuned outside the qubit band, the same loss becomes useful. It can drain excitations from the qubit and serve as a reset channel. Both regimes are examined through circuit simulations in Fig.\ \ref{fig:FigS2}(a), where ``Length" denotes the coupling-feedline length and ``Location" denotes the position of the resonator and qubit along that feedline.

Taking the $13$-order HPF and LPF as examples, we first simulate the bare filters without the resonator or qubit and vary the feedline length. As shown in Figs.\ \ref{fig:FigS2}(b) and (d), the dissipation-mode frequency scales roughly inversely with the length, whereas the cutoff frequency is almost unaffected by this length. The dissipation mode can therefore be moved away from the qubit band by choosing the feedline length. In practice, a length of $4$--$6$~mm is appropriate for the HPF and $5$--$10$~mm for the LPF. Beyond about $10$~mm, the higher-order harmonics of the dissipation mode on the feedline move close to the qubit band, and their influence needs to be taken into account.

Whether the mode can act as a reset channel further depends on the coupling location of the resonator along the feedline. The reason is the voltage profile of the mode, which fixes how strongly the resonator and qubit at a given location couples to it. For the $13$-order HPF and LPF, we fix the coupling-feedline length at $5$~mm according to the above analysis, and vary the coupling location of the resonator and qubit. As shown in Figs.\ \ref{fig:FigS2}(c) and (e), a qubit at the center of the feedline sits at the voltage node\cite{sunada2022fast} of the dissipation mode. Its coupling to the mode vanishes and it sees zero dissipation, so the reset channel is unavailable at this point. To exploit the reset channel, the resonator and qubit should be placed away from the node, where the normalized qubit dissipation $\Gamma_q$ rises accordingly.

\subsection{One-section versus two-section filters}\label{SecS2.2}

\begin{figure}[htb]
\includegraphics{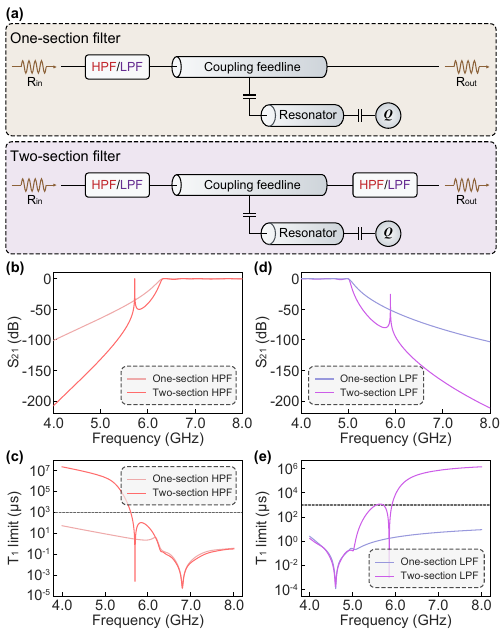}
\caption{\textbf{One-section versus two-section filters.} (a) Schematics of the one-section (top) and two-section (bottom) filters. (b)(d) Simulated transmission spectra of the $13$-order one-section and two-section HPF and LPF. (c)(e) Corresponding Purcell-limited relaxation times; the dashed line marks the $1$~ms reference level.}
\label{fig:FigS3}
\end{figure}

In our filters, we adopt a two-section design rather than a one-section design\cite{park2024characterization}, so that the qubit is protected on both sides. In the one-section design, the filter consists of compact lumped elements, which are not convenient for coupling a readout resonator. The resonator therefore has to couple to the transmission line on the side without a filter. Through the resonator the qubit can radiate to that side directly, which leaves it unprotected there and shortens its relaxation times. The filter can be placed at either side, and both choices give the qubit the same external dissipation, so we show only one representative case. The two-section design instead puts a filter on both sides and strongly suppresses this dissipative path.

We compare the one-section and two-section designs for $13$-order filters. Their transmission spectra are obtained from circuit simulations and shown in Fig.\ \ref{fig:FigS3}(b) and (d). In both cases, the resonator is kept at the same location relative to the filter and given the same external quality factor. We then extract the corresponding $T_1$ limits using the black-box method of the main text [Eq.\ (4)]. As shown in Fig.\ \ref{fig:FigS3}(c) and (e), at the qubit frequency the two-section design reaches a much higher $T_1$ limit than the one-section design, which agrees with our expectation.

\subsection{Asymmetric edge-pass filters}\label{SecS2.3}

\begin{figure}[htb]
\includegraphics{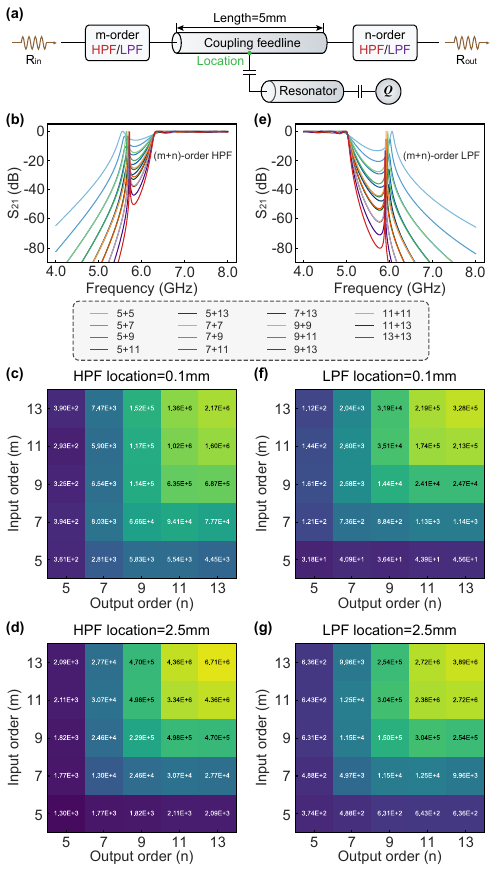}
\caption{\textbf{Asymmetric edge-pass filters.} (a) Schematic of the asymmetric filter, with an $m$-order section at the input and an $n$-order section at the output and the feedline length fixed at $5$~mm. (b)(e) Simulated transmission spectra of the asymmetric HPF and LPF for $m,n=5\sim13$. (c)(f) Purcell-limited relaxation times at the qubit frequency of the HPF and LPF with the resonator and qubit at location $=0.1$~mm. The colorbar represents the simulated qubit $T_1$ limit in units of $\mu$s. (d)(g) The same at location $=2.5$~mm.}
\label{fig:FigS4}
\end{figure}

Finally, we consider asymmetric filters, in which the input and output sides carry sections of different order. Such a design adds a degree of freedom to the layout, since the two sides need not to be identical. In the configuration of Fig.\ \ref{fig:FigS4}(a), an $m$-order section is placed at the input side and an $n$-order section at the output side, with the coupling feedline length fixed at $5$~mm.

The transmission spectra for $m,n=5\sim13$ are shown in Fig.\ \ref{fig:FigS4}(b) and (e). The dissipation-mode frequency is set mainly by the feedline length and remains nearly the same across the $(m,n)$ combinations, whereas a larger total order $m+n$ steepens the roll-off and deepens the stopband. The corresponding simulated qubit $T_1$ limits (in $\mu$s) at the qubit frequency are tabulated for the two filters at two locations of the resonator and qubit. Throughout these simulations, the resonator frequency and external quality factor are kept nearly the same. For the HPF, the limits are given in Fig.\ \ref{fig:FigS4}(c) and (d) at location $=0.1$~mm and $=2.5$~mm, respectively. For the LPF, they are given in Fig.\ \ref{fig:FigS4}(f) and (g) at the same two locations. As an example, the entry with input side $m=5$ and output side $n=5$ of the HPF at location $=0.1$~mm reads $3.61\times10^2~\mu$s, the Purcell-limited relaxation time at the qubit frequency ($5$~GHz). The remaining entries follow the same convention, with the LPF qubit frequency taken at $6.5$~GHz. Across all four maps, the $T_1$ limit spans several orders of magnitude and grows as the total order $m+n$ increases, consistent with the deepening stopband of the transmission spectra. Comparing the two locations, the values at location $=2.5$~mm are generally higher than those at location $=0.1$~mm for both filters. These maps provide a reference for choosing the section orders and the resonator-qubit location that meet a target $T_1$ limit.

\makeatletter
\renewcommand\@biblabel[1]{[#1.]}
\makeatother

%

\bigskip